\documentclass{iau}
\usepackage{graphicx}
\usepackage{subfigure}

\title[Str\"omgren photometry and  spectroscopy  
] 
{ Characterizing the variability of Melotte~111~AV~1224: a new variable star in the Coma Berenices open cluster}

\author[L. Fox-Machado et al.]   
{L. Fox-Machado, R. Michel, M. Alvarez and J.H. Pe\~na}

\affiliation{Instituto de Astronom\'{\i}a, Universidad Nacional Aut\'onoma de M\'exico\\ e-mails: {\tt lfox@astrosen.unam.mx, rmm@astrosen.unam.mx,  alvarez@astrosen.unam.mx,jhpena@astroscu.unam.mx  }\\
}

\pubyear{2014}
\volume{301}  
\pagerange{1--2}
\jname{Precision Asteroseismology}
\editors{J.A. Guzik, W.J. Chaplin, G. Handler \& A. Pigulski, eds.}
\begin{document}

\maketitle

\begin{abstract}
A search for new pulsating stars in the Coma Berenices open cluster was carried out.
As a result of this search, the cluster member  
 Melotte~111~AV~1224 presented clear indications of photometric
variability.
In order to determine its physical parameters, Str\"omgren standard indices
  and low-resolution spectra were acquired.
 In this work, we present the preliminary results of these observations.
\end{abstract}

\firstsection 
\section{Introduction}

Coma Berenices  (Melotte~111, RA\,$=12^{\rm h}$23$^{\rm m}$, DEC\,$=+26^{\rm o}$00$^{\prime}$, J2000.0) 
is the second closest open cluster to the Sun. The \textit{Hipparcos} distance of Melotte 111 is
  $d=89.0 \pm 2.1$~pc (\cite[van Leeuwen 1999]{van}), in  agreement with older ground-based estimates 
(e.g. $85.4 \pm 4.9$~pc, \cite[Nicolet 1981]{nicolet}).  The metallicity of the  cluster has been derived by several authors.  
For example, \cite[Cayrel de Strobel (1990)]{cayrel} determined [Fe/H] $=-0.065 \pm 0.021$~dex, 
whereas \cite[Friel \& Boesgaard (1992)]{friel} found [Fe/H] $=0.052 \pm 0.047$~dex. The age of the cluster is 
estimated between 400 and 500 Myr (\cite[Bounatiro \& Arimoto 1993]{boun}). As its physical parameters
are well constrained, the variability studies in Melotte~111  are very important. We have carried out
a search for new pulsating stars in the direction of Melotte 111. As a result of this search,
the star Melotte~111~AV~1224 was found to be a new variable star. This
star was originally designated AV~1224 in the
astrometric catalogue for the area of Coma Berenices (\cite[Abad \& Vicente 1999]{abad}).
It is listed as a cluster member in the Simbad database.
This work presents preliminary results aimed at characterizing the variability of this target.

\section{Observations, data reduction and conclusion}
\begin{figure}
 \subfigure[]{\includegraphics[width=0.53\textwidth]{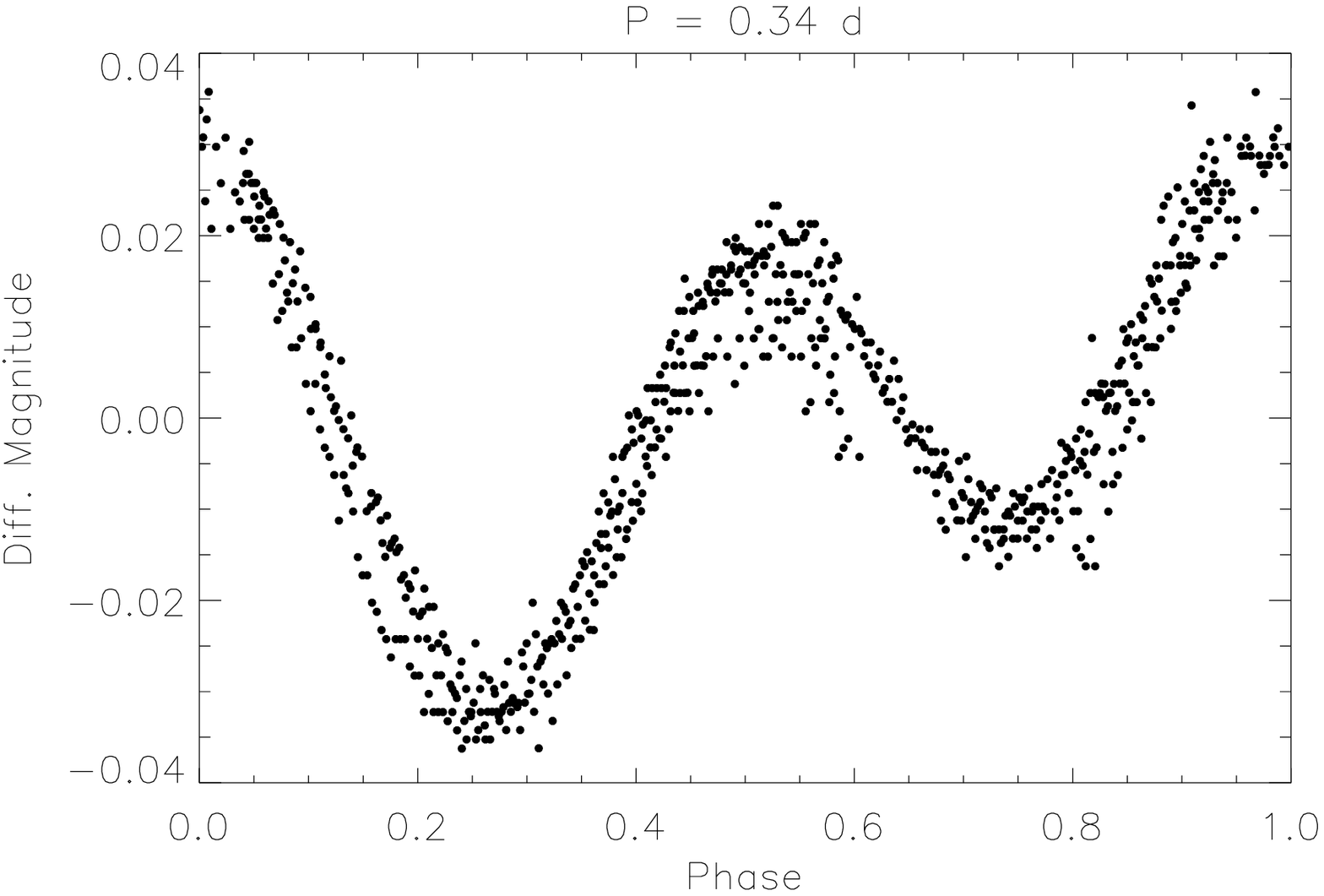}}
 \subfigure[]{\includegraphics[width=0.47\textwidth]{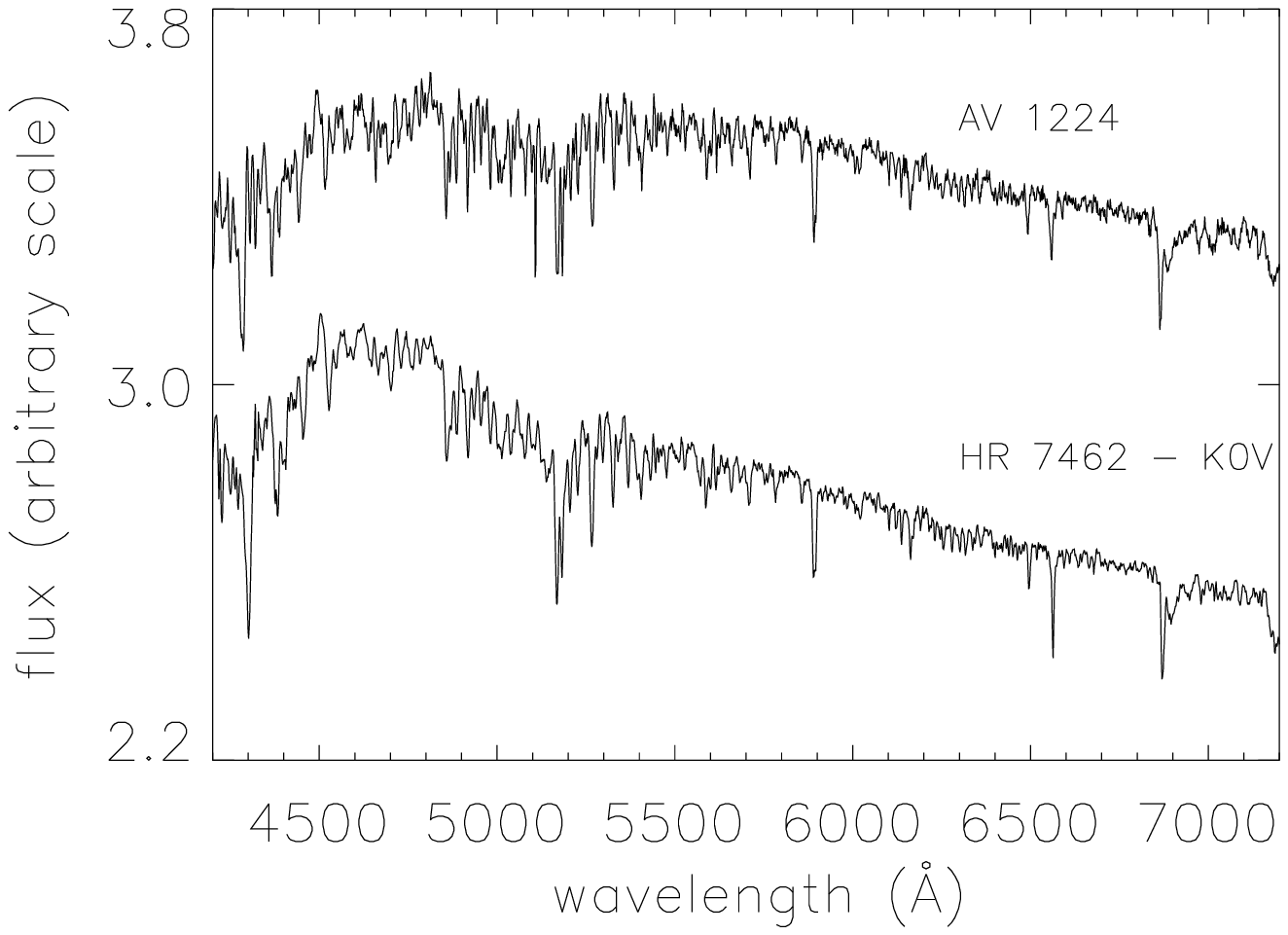}}
 \caption{(a) Light curve of AV~1224 phased with the period of 0.34~d. (b) Spectrum of AV~1224 and HR~7462, a
star of similar spectral type.}
\label{fig1}
 \end{figure}

The CCD  observations of the Melotte~111 open cluster have been made with the 0.84-m
f/15 Ritchey-Chr\'etien telescope at OAN-SPM observatory,  
during  ten consecutive nights, between April 11 and 20, 2009. The 
telescope hosted the filter-wheel `Mexman' with the Marconi (E2V) CCD camera,
which has a 2048 $\times$ 2048 pixels array, with a pixel size of 15 $\times$ 15 $\mu$m$^{2}$.
The typical field-of-view in this configuration amounts to 7$^{\prime}$ $\times$ 7$^{\prime}$.
The observations were obtained through a Johnson $V$ filter. 
The usual calibration procedures for CCD photometry have been
carried out during  our observing run. Sky flat fields, bias and dark exposures were taken every night.
The resulting light curve is not sinusoidal, but is strictly periodic; the frequency spectrum reveals two peaks,
$2f_{1} \sim 5.8$~d$^{-1}$, $A =$ 23.2~mmag  and
$f_{1} \sim 2.9$~d$^{-1}$, $A =$ 10.6~mmag. The light curve of AV~1224 phased with its main period, is shown in Fig.~\ref{fig1}a.  
We have also derived the following indices in the Str\"omgren system for AV~1224: 
$V=13.709$, $(b-y)=0.526$,  $m_1=0.276$, and $c_1=0.320$. 

Spectroscopic observations of the star were conducted with the 2.12-m
telescope of the OAN-SPM observatory in June 2011. We used  the same equipment as explained by \cite[Baran et al. (2011)]{baran}.
In particular, we used 
Boller \& Chivens spectrograph installed in the Cas\-se\-grain focus of the telescope. 
The 400 lines/mm grating with a blaze angle of 4.18$^{\circ}$ was used.
The grating angle was set to 7$^{\circ}$ to cover wavelength range from 4000\,{\AA} to 7500\,{\AA}.
A 2048$\times$2048 E2V CCD camera
was used in the observations. 
The typical resolution of the recorded spectra is 8\,{\AA} and the dispersion amounts to
1.8\,{\AA} per pixel. The reduction procedure was performed with the standard routines
of the IRAF package.  Fig.~\ref{fig1}(b) shows the reduced spectrum of AV~1224 and, for  comparison, the 
spectrum of a standard
star of spectral type K0\,V  taken on the same night.
Considering the Str\"omgren indices and the stellar spectrum, the variability of AV~1224 due
to pulsations can be ruled out. Its light curve 
 resembles rather those observed in W Ursae Majoris-type variables (W\,UMa), also called EW stars.
 As it is known, the components of W UMa systems are
in contact and are
main-sequence stars of nearly the same spectral type, from
around middle A to early K. Their orbital periods range from 0.2 to 1.4 days. An in-deep analysis of these
observations will be given elsewhere.

The authors acknowledge the financial support from the UNAM through grant
PAPIIT IN104612 and from the IAU.

\end{document}